\documentclass[letterpaper,english,showpacs,reprint]{revtex4-1}
\usepackage[T1]{fontenc}
\usepackage[latin9]{inputenc}
\usepackage{amsmath}
\usepackage{amssymb}
\usepackage{graphicx}
\usepackage{esint}

\makeatletter
\usepackage[colorlinks=true, linkcolor=blue,citecolor=blue,urlcolor=blue]{hyperref}

\makeatother

\usepackage{babel}
\begin{document}

\title{Interference and nonlinear properties of four-wave-mixing resonances
in thermal vapor: Analytical results and experimental verification}

\author{Micha\l{} Parniak}

\email{michal.parniak@fuw.edu.pl}

\selectlanguage{english}%

\affiliation{Institute of Experimental Physics, University of Warsaw, Pasteura
5, 02-093 Warsaw, Poland}

\author{Wojciech Wasilewski}

\affiliation{Institute of Experimental Physics, University of Warsaw, Pasteura
5, 02-093 Warsaw, Poland}
\begin{abstract}
We develop a model to calculate nonlinear polarization in a nondegenerate
four-wave mixing in diamond configuration which includes the effects
of hyperfine structure and Doppler broadening. We verify the model against
the experiment with $5^{2}S_{1/2}$, $5^{2}P_{3/2}$, $5^{2}D_{3/2}$
and $5^{2}P_{1/2}$ levels of rubidium 85. Treating the multilevel atomic system as a combination of many four-level systems we are able to express the nonlinear susceptibility of a thermal ensemble
in a low-intensity regime in terms of Voigt-type profiles and obtain an excellent conformity of theory and experiment within this complex system. The agreement is also satisfactory
at high intensity and the analytical model correctly predicts the
positions and shapes of resonances. Our results elucidate
the physics of coherent interaction of light with atoms involving
higher excited levels in vapors at room temperature, which is used
in  an increasing range of applications.
\end{abstract}


\maketitle

\section{Introduction}

Coherent interactions of light and atomic vapors involving higher
excited states have attracted much attention recently. A seminal work
by Peyronel\emph{ et al.} \cite{Peyronel2012} demonstrated
electromagnetically induced transparency resonance with extreme single-photon
sensitivity due to utilization of Rydberg levels. A host of other
works explore the use of multi-photon transitions involving higher
states in the context of non-linear optics and quantum information. 

\begin{figure}
\includegraphics[scale=0.3,bb=0bp 390bp 570bp 842bp,clip]{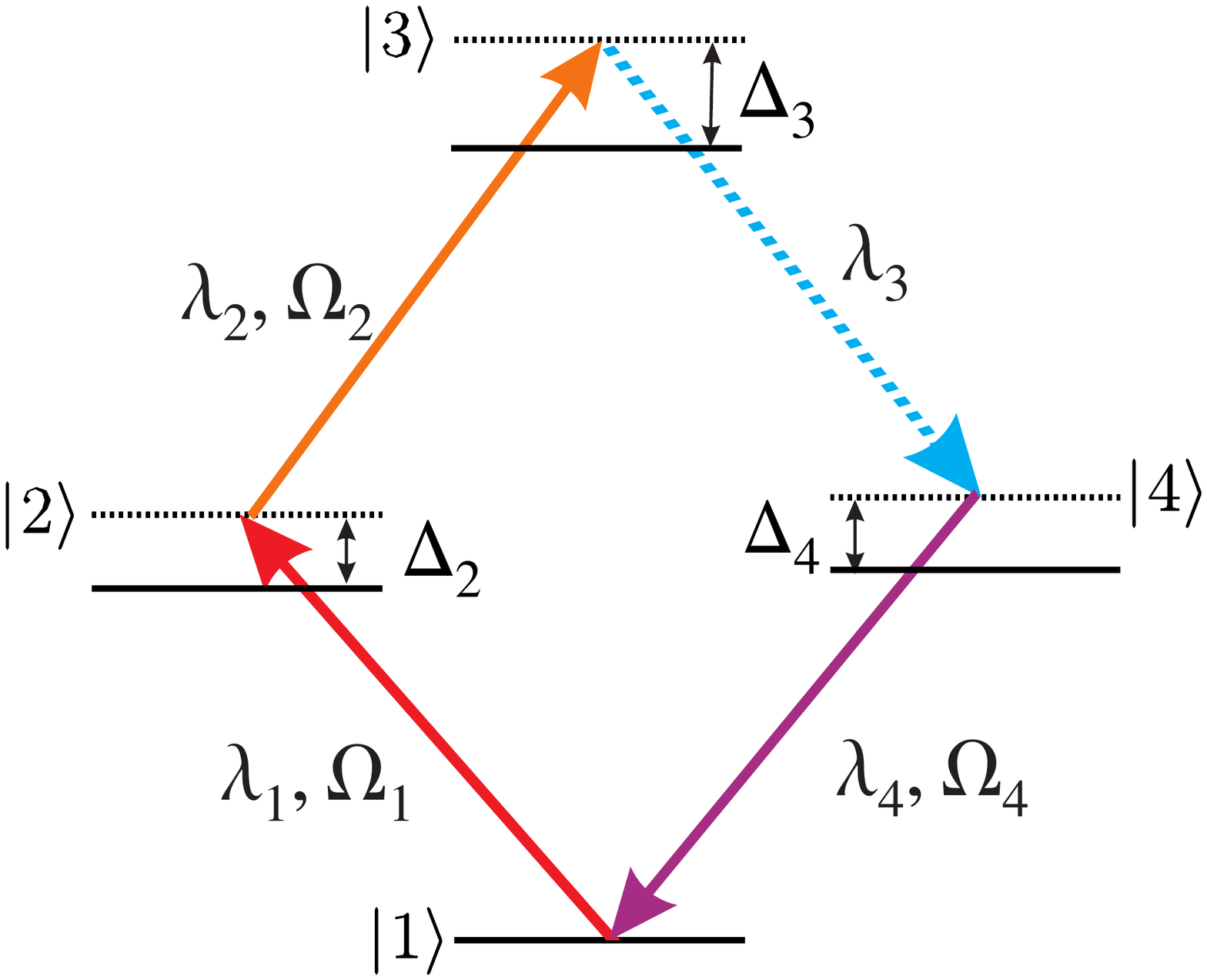}

\protect\caption{(Color online) The diamond scheme of atomic levels and transitions. Rubidium 85 is
used to generate light at the wavelength $\lambda_{3}=762$~nm via
4WM of incident light fields at the wavelengths $\lambda_{1}=780$~nm, $\lambda_{2}=776$~nm and $\lambda_{4}=795$~nm, and Rabi frequencies
$\Omega_{1}$, $\Omega_{2}$ and $\Omega_{4}$. Levels $|1\rangle$
to $|4\rangle$ correspond to $5^{2}S_{1/2}$, $5^{2}P_{3/2}$, $5^{2}D_{3/2}$
and $5^{2}P_{1/2}$ states of rubidium 85, respectively. The optical
frequencies are parametrized by two one-photon detunings $\Delta_{2}$
and $\Delta_{4}$ and a two-photon detuning $\Delta_{3}$. Note that the single photon detuning of the $\Omega_2$ field can be written as $\Delta_3-\Delta_2$.}
\label{fig:lsch}
\end{figure}

The so-called diamond configuration of atomic levels, as sketched
in Fig. \ref{fig:lsch}, is frequently used. Applications include:
coherent interaction of atoms with ultrashort laser pulses,
useful for interferometric measurements \cite{Clow2010}, and coherent
control \cite{Lee2013}. In the continuous-wave regime, atoms in diamond
configuration demonstrate high resonant non-linearities that enable
generation of coherent blue and infrared light \cite{Sell2014,Meijer2006,Zibrov2002,Akulshin2009a}
via four-wave mixing (4WM), as well as strong phase-dependent response \cite{Morigi2002}. This up-conversion scheme also enables
coherent transfer of light angular momentum \cite{Walker2012}.

Atoms in diamond configuration have been successfully used as a nonlinear
frequency conversion medium \cite{Chaneliere2006,Donvalkar2014,Radnaev2010a,Jen2010},
applicable to light at single photon level. Both spectral \cite{Willis2009a,Becerra2008}
and temporal \cite{Becerra2010} properties of the diamond configuration
have been studied. 

When the diamond configuration involved a Rydberg state, 4WM has also
been observed \cite{Kolle2012} and coherent revival effects have
been seen even in a warm ensemble of atoms \cite{Huber2014a}. A similar
configuration is also proposed to be a room-temperature single photon
source \cite{Muller2013}.

Finally, 4WM in diamond configuration has been used as a source of
single photon pairs \cite{Willis2011,Zhang2014} with well-defined
temporal properties \cite{Srivathsan2013a} which, thanks to proper
postselection, can be perfectly matched to absorption by single atoms
\cite{Gulati2014a}.

In most of the afore-mentioned works a maximally five-level atom model
was used to model the experiment. 
Thereby the interferences between different paths through rich hyperfine 
structure of actual atoms were neglected.
The hyperfine structure inevitably influences any room-temperature
experiments where different transitions are driven in classes of atoms
depending on their velocity. The susceptibility of a room-temperature
ensemble is a convolution of respective cold-atomic function in a
multidimensional space of frequencies of the driving fields with a
Gaussian corresponding to thermal velocity distribution. This particular
convolution reshapes resonances and forms band structures which extend
in certain directions while they stay sub-Doppler in others. 
It is therefore very important to take into account contributions of all hyperfine states
or, in other words, a multitude of paths through intermediate levels.

In this paper we develop simple theoretical tools for predicting the
influence of this effect on a coherent process in which it is crucial to consider interference
between contributions of different atomic states. We introduce a formalism that enables us to treat the realistic atom as a combination of many four-level atoms. This can be understood as an interference of many possible paths of 4WM. For weak driving fields the final result is a superposition of Voigt-type
profiles, and a relatively simple recipe is provided for higher
intensities. The simplicity worked out enables us to explain the intricacies
of interference between contributions of various hyperfine sublevels,
taking into account the Doppler broadening in a precise and exact
way.

We verify our results against experiment in which we measure the intensity
of 4WM in diamond configuration in warm rubidium 85 vapor as a function
of laser frequencies. The comparison covers both low and high intensity
regimes. We obtain good conformity of theory and experiment despite complexity of the system.

The effects we describe and explain herein are especially important
to studying phase-dependent interactions \cite{Morigi2002,Lee2013,Clow2010},
to optimizing single-photon generation setups \cite{Srivathsan2013a,Zhang2014},
and to explaining 4WM spectra observed in many experiments \cite{Willis2009a,Becerra2008,Akulshin2009a}
or possible magnetic field effects. 

This paper is organized as follows. In Sec. \ref{sec:Theory} we
introduce a theoretical description of 4WM in an inhomogeneously broadened
multi-level atomic medium. In Sec. \ref{sec:Experiment} we describe
the details of our 4WM experiment in rubidium vapors. Section \ref{sec:Results}
presents a comparison of experimental and theoretical data. Sec. \ref{sec:Conclusions} concludes the paper.

\section{Theory\label{sec:Theory}}

In this section we calculate the intensity of the light generated
by 4WM in diamond configuration depicted in Fig. \ref{fig:lsch}.
We start with calculating the optical coherence in a four-level atom
model. Next we take into account the hyperfine structure, and finally
the Doppler broadening.

\subsection{Nonlinear polarization of a four-level atom}

In the experiment we observe the intensity of light generated at the
transition between levels $|3\rangle$ and $|4\rangle$ at $\lambda_{3}=762$
nm via 4WM. The starting point for the theoretical treatment is the
calculation of steady-state optical coherence $\rho_{43}$ in a four-level
atom in diamond configuration. For the sake of simplicity we assume
that the intensity of the incident laser light remains constant along
the atomic medium. When the geometry of the laser beams enables perfect
wave-vector matching, the amplitude of the emitted light wave $E_{3}$
is proportional to the optical polarization $P_{3}$, which, in turn,
is proportional to the density of atoms $n$, the dipole moment of
the transition $\mu_{43}^{*}$, and the optical coherence 
\begin{equation}
E_{3}\propto P_{3}=n\mu_{43}^{*}\rho_{43}+\mathrm{c.c.}\label{eq:P3=00003Dnmurho}
\end{equation}

The steady-state coherence $\rho_{43}$ can be calculated from the
Liouville equation with relaxation, constructed as in Ref. \cite{Kolle2012}.
For low light intensities, we may use the lowest nonvanishing order
of the perturbative solution and obtain a compact formula for the
optical coherence:

\begin{equation}
\rho_{43}=\frac{{\Omega_{1}}\Omega_{2}\Omega_{4}^{*}}{8\tilde{\Delta}_{2}\tilde{\Delta}_{3}\tilde{\Delta}_{4}^{*}},\label{eq:coherence-p}
\end{equation}

where $\tilde{\Delta}_{j}=\Delta_{j}+i{\Gamma_{j}}/2$ is the complex
detuning of field from level $|j\rangle$, $\Gamma_{j}$ is the decay rate of level $|j\rangle$, $\Omega_{k}$ is
the Rabi frequency of field $E_k$ coupled to $|k\rangle-|k+1\rangle$ transition for $k=1,2$, and $\Omega_4$ is the Rabi frequency of field $E_4$ coupled to $|1\rangle-|4\rangle$ transition, in accordance with Fig. \ref{fig:lsch}. For rubidium 85, the decay rates are $\Gamma_2=6.1\times2\pi$~MHz, $\Gamma_3=0.66\times2\pi$~MHz and $\Gamma_4=5.7\times2\pi$~MHz. This leads
to the following expression for the optical polarization:
\begin{equation}
P_{3}=n\frac{{\mu_{12}}\mu_{23}\mu_{43}^{*}\mu_{14}^{*}}{8\hbar^{3}\tilde{\Delta}_{2}\tilde{\Delta}_{3}\tilde{\Delta}_{4}^{*}}E_{1}E_{2}E_{4}^{*}+\mathrm{c.c.}.\label{eq:P3=00003DEmuEmuEmu/DDD}
\end{equation}

For driving light intensities exceeding saturation $\Omega\gtrsim\Gamma$
we need to solve the Liouville equation exactly. A computer algebra
system provides an exact solution in the form of a rational function
of complex detunings $\tilde{\Delta}_{i}$ and Rabi frequencies $\Omega_{i}$.
We state this solution in a general way:

\begin{equation}
\rho_{43}=\rho_{43}(\tilde{\Delta}_{2},\tilde{\Delta}_{3},\tilde{\Delta}_{4}^{*},\Omega_{1},\Omega_{2},\Omega_{4}^{*}).\label{eq:coherence-e}
\end{equation}

\begin{figure}
\includegraphics[scale=0.58]{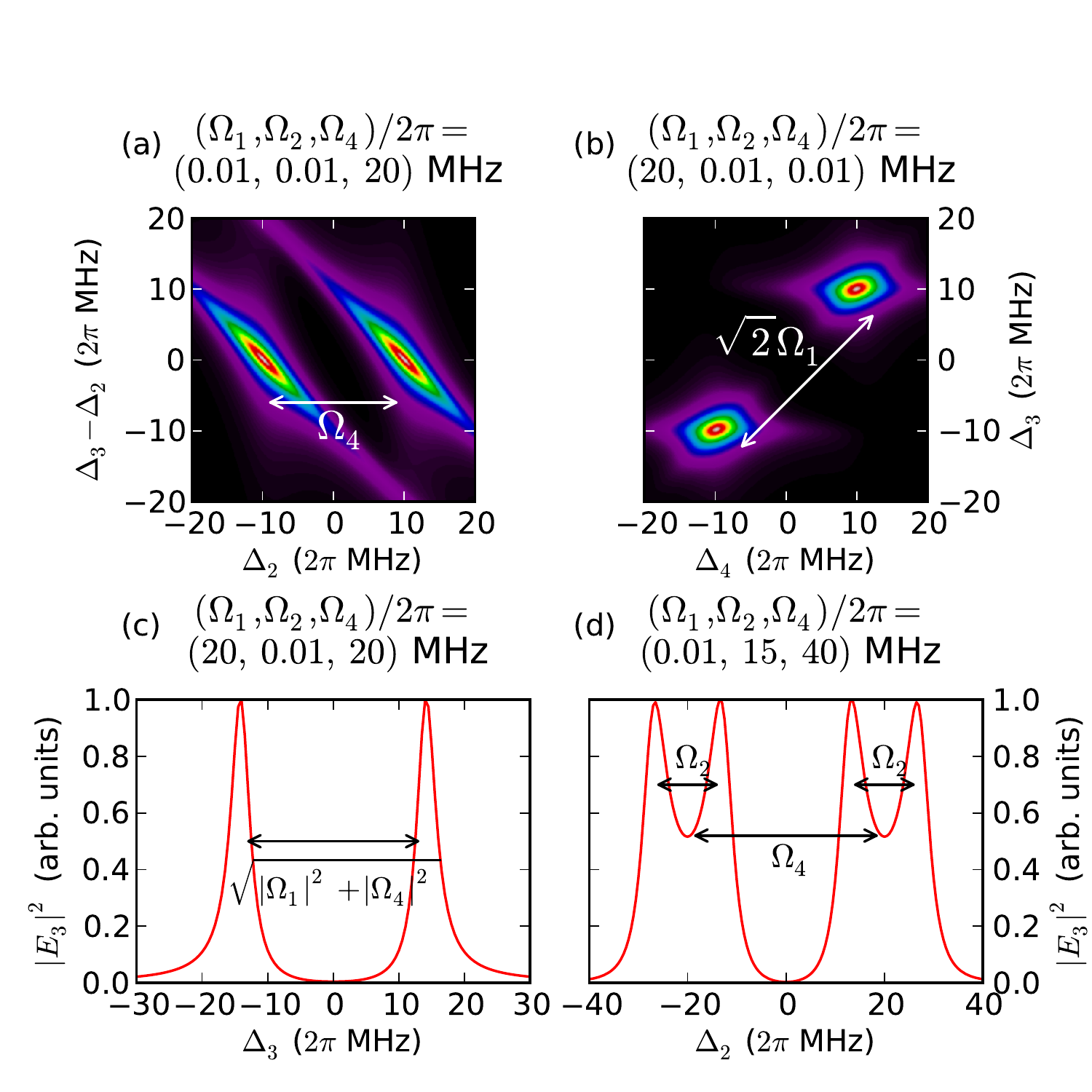}\protect\caption{(Color online) Intensity of the 4WM signal $|E_3|^2$: (a) the detuning of strong $\Omega_4$ driving field is kept constant ($\Delta_4=0$) while the frequencies of the $\Omega_1$ field ($\Delta_2$ detuning) and $\Omega_2$ field ($\Delta_3-\Delta_2$ detuning) are swept to probe the splitting, (b) the detuning of $\Omega_1$ field is kept constant  ($\Delta_2$=0) while the frequencies of $\Omega_2$ ($\Delta_3$) and $\Omega_4$ ($\Delta_4$) fields are swept, (c) two strong driving fields ($\Omega_1$ and $\Omega_4$) of constant frequency ($\Delta_2$=0 and $\Delta_4$=0) induce a splitting of $\sqrt{|\Omega_1|^2+|\Omega_4|^2}$, which is probed by a weak $\Omega_2$ field detuned by $\Delta_3$ from resonance, (d) strong $\Omega_2$ field of constant single-photon detuning from the $|2\rangle-|3\rangle$ resonance ($\Delta_3=\Delta_2$) and strong $\Omega_4$ field ($\Delta_4=0$) causes four peaks to appear when $\Delta_2$ detuning is varied.  \label{fig:cold}}

\end{figure}

In Fig. \ref{fig:cold}, we plot the 4WM signal intensity $|E_{3}|^{2}$
in various situations, calculated using the
solution (\ref{eq:coherence-e}) in Eq.~(\ref{eq:P3=00003Dnmurho}),
to illustrate an already complex behavior of the system. For high
Rabi frequencies, i.e. $\Omega\gtrsim\Gamma$, the system reveals Autler-Townes splitting \cite{Becerra2008,Autler1955} observed in the emission of the signal field.
To directly observe the splitting, we set one driving field to be strong [$\Omega_4/2\pi=20 $~MHz in Fig \ref{fig:cold}(a) and $\Omega_1/2\pi=20$~MHz in Fig. \ref{fig:cold}(b)], while frequencies of the remaining weak fields are swept to probe the structure. In both cases, we see two resonances of the signal field, and their separation is proportional to the strong field Rabi frequency.
Even more complex behavior, that can be understood as double dressing of atomic states \cite{Wei1998}, reveals itself when two of the driving fields are strong. In Fig. \ref{fig:cold}(c) we set $\Omega_1/2\pi=20$~MHz and $\Omega_4/2\pi=20$~MHz. When the two-photon detuning $\Delta_3$ is changed, we observe two resonances separated by $\sqrt{|\Omega_1|^2 + |\Omega_4|^2}$. Finally, following \cite{Jen2010}, we set $\Omega_2/2\pi=15$~MHz and $\Omega_4/2\pi=40$~MHz and observe four distinct resonances of four-wave mixing in Fig. \ref{fig:cold}(d), arising due to double Autler-Townes splitting.

The above solutions can be readily applied to describe 4WM in a cold
ensemble of atoms or a single atom where the hyperfine structure of
each of the intermediate levels is fully resolved. Even though these solutions can already be quite complex, they do not provide significant insight into the behavior of the Doppler broadened medium, where direct observation of Autler-Townes splitting becomes harder.

\subsection{Hyperfine structure: Interference of paths}

In order to fully account for the hyperfine structure of a real atom,
we would have to solve the Liouville equation for an excessively large
density matrix. Later in this paper we deal with an experimental case where the equation would have to take into
account at least 64 atomic states. Solution of the resulting steady-state
Liouville equation for a range of parameters, such as detunings would
be hard to compute.

Instead, we propose an alternative solution to this problem that can be computed much more rapidly. For the sake of simplicity,
we assume that all light fields, both the driving ones and the signal
one, have a fixed polarization parallel to the quantization axis $x$,
as this is our experimental situation.

In our method, we treat the full system as a set of many four-level
systems. Each of these can be represented as a certain path $\mathcal{{P}}=\{|F_1\, m_{F_{1}}\rangle,|F_{2}\, m_{F_{2}}\rangle,|F_{3}\, m_{F_{3}}\rangle,|F_4\, m_{F_{4}}\rangle,|F_1\, m_{F_{1}}\rangle\}$,
through intermediate levels, where $F_{i}$ is the total angular momentum
of the state $|i\rangle$, and $m_{F_{i}}$ is its projection. As a path $\mathcal{P}$ we understand a list of five atomic states. Each state belongs to the hyperfine manifold of appropriate level $|1\rangle$ to $|4\rangle$ and is characterized by a specific $F$ and $m_F$ for a given path. Each
path can be characterized by different dipole moments $\mu_{ij}(\mathcal{P})$,
and different hyperfine detunings $\Delta_{i}(\mathcal{P})$. Angular
momenta for all of the excited levels $|2\rangle$, $|3\rangle$, and
$|4\rangle$ vary through all possible values, but for the ground
state $|1\rangle$, which is both the initial and the final state of the path, we choose only one ground-state angular momentum
$F_{1}$ and vary only its projection $m_{F_{1}}$. This is due to
the fact that the ground-state hyperfine manifold is fully resolved
in our experimental case. 

The dipole moments $\mu_{ij}(\mathcal{P})=\langle J_{i}\mathcal{\, I}\, F_{i}\, m_{F_{i}}|\hat{{\mu}}|J_{j}\,\mathcal{I\,}F_{j}\, m_{F_{j}}\rangle$,
where $\hat{{\mu}}$ is the dipole moment operator, $\mathcal{I}$
is the nuclear spin, and $J_{i}$ is the total angular momentum of
the electron on the level $|i\rangle$, are calculated using known
values of reduced dipole moments $\mu_{ij}=\langle J_{i}||\hat{{\mu}}||J_{j}\rangle$
according to the formulas given in Ref.~\cite{Steck2012}. The complex
detunings $\tilde{\Delta}_{i}(\mathcal{P})$ calculated from respective lines are affected by hyperfine shifts in the following
way: 
\begin{equation}
\tilde{\Delta}_{i}(\mathcal{P})=\tilde{{\Delta}}_{i}+\Delta F_{i}-\Delta F_{1},
\end{equation}
where $\Delta F_{i}$ is the shift of a given $F_{i}$ from the centroid
of the manifold $|i\rangle$. It follows that the detunings $\tilde{\Delta}_{i}$
are measured to the manifold centroids.

In our experimental case 61 paths through intermediate states contribute
to the optical polarization $P_{3}$. For each path $\mathcal{{P}}$
we calculate its contribution $P_{3}(\mathcal{P})$ by inserting solution
(\ref{eq:coherence-e}) into Eq.~(\ref{eq:P3=00003Dnmurho}). In
total, we obtain:
\begin{equation}
P_{3}=\sum_{\mathcal{P}}n(\mathcal{P})\mu_{43}^{*}(\mathcal{P})\rho_{43}(\mathcal{P})+\mathrm{c.c.},\label{eq:P3=00003Dsumpaths}
\end{equation}
where $n(\mathcal{P})$ is the density of the atoms in the ground
state sublevel $F_{1},\, m_{F_{1}}$of a particular path $\mathcal{P}$,
while $\rho_{43}(\mathcal{P})$ is calculated taking into account
hyperfine shifts and dipole moments corresponding to the given path
$\mathcal{P}$. The above equation expresses an approximation neglecting
coherent interplay of different paths in a single atom.

However, such interplays may only matter when two paths share common
levels and some coherences oscillate at similar frequencies. In our
experimental situation, all of the optical fields are polarized linearly
in the same direction $x$, which we take as quantization axis. In
this case, all $m_{F_{i}}$ in each path are the same due to selection
rules for dipole moments. Consequently if two distinct paths share
a level $|i\rangle$, then they must have a different total angular
momentum $F_{j}$ at some other level $|j\rangle$. The natural oscillation
frequencies of coherences for these paths differ by the hyperfine
splitting. As the Rabi frequencies in our experiment are smaller than
the hyperfine splittings, we conjecture such paths will not interplay
coherently. This is verified in the experiment.

Strong driving beams significantly redistribute the atoms among the
ground-state sublevels $F_{1},m_{F_{1}}$ altering $n(\mathcal{P})$.
We find the steady state of a set of rate equations to determine $n(\mathcal{P})$
and thus the relative contributions of paths. 

To conclude this section, let us note that within the limit of low
intensities of driving fields $\Omega\ll\Gamma$, the total
polarization can be calculated exactly using perturbation calculus.
In the lowest nonvanishing order, Eq.~(\ref{eq:P3=00003DEmuEmuEmu/DDD})
can be inserted into Eq.~(\ref{eq:P3=00003Dsumpaths}) leading to
a simple result:

\begin{equation}
P_{3}=\epsilon_{0}E_{1}E_{2}E_{4}^{*}\sum_{\mathcal{P}}\chi(\mathcal{P})+\mathrm{c.c.},\label{eq:P3=00003DEEEnchi}
\end{equation}

where we identified the susceptibility $\chi(\mathcal{P})$ of the
system:

\begin{equation}
\chi(\mathcal{P})=n(\mathcal{P})\frac{\mu_{12}(\mathcal{P})\mu_{23}(\mathcal{P}){\mu_{43}^{*}}(\mathcal{P})\mu_{14}^{*}(\mathcal{P})}{8\epsilon_{0}\hbar^{3}\tilde{\Delta}_{2}(\mathcal{{P}})\tilde{{\Delta}}_{3}(\mathcal{{P}})\tilde{{\Delta}}_{4}^{*}(\mathcal{{P}})}.\label{eq:chi=00003Dmu/Delta}
\end{equation}

The above expression approximates the rough features of full non-perturbative
solution quite well and we find it worthwhile to track the contributions
of various hyperfine components to the final result. We list them
in Table~\ref{tab:tab} for the 4WM process we study in experiment.

\begin{table}
\begin{tabular}{|c|c|c|c|c|c|c|c|c|c|c|c|c|}
\hline 
$\chi(\mathcal{P})$ & \multicolumn{4}{c|}{$F_{2}=2$} & \multicolumn{4}{c|}{$F_{2}=3$} & \multicolumn{4}{c|}{$F_{2}=4$}\tabularnewline
\hline 
\hline 
$|m_{F_{1}}|$ & 3 & 2 & 1 & 0 & 3 & 2 & 1 & 0 & 3 & 2 & 1 & 0\tabularnewline
\hline 
$F_{4}=2$ & 0 & -40 & 32 & 72 & 0 & 140 & 56 & 0 & 0 & 180 & 360 & 432\tabularnewline
\hline 
$F_{4}=3$ & 0 & 80 & 32 & 0 & 360 & 0 & -21 & 0 & 216 & 144 & 45 & 0\tabularnewline
\hline 
\end{tabular}

\protect\caption{Theoretical relative amplitudes of 4WM resonances in rubidium 85 ($F_3$ ground state) parametrized
by spin of $|2\rangle$ and $|4\rangle$ states $F_{2}$ and $F_{4}$
for each spin projection of the ground state $m_{F_{1}}$. The result is summed over the spin of highest excited state $F_{3}=1\ldots4$ due to negligible
hyperfine splitting.}
\label{tab:tab}
\end{table}

\subsection{Doppler broadening}

The final step in the construction of our model consists in averaging
the contributions of atoms moving with various velocities $v$ along
almost parallel laser beams. We take the one-dimensional Maxwell-Boltzmann
velocity distribution $g(v)=\sqrt{\frac{{m}}{2\pi k_{B}T}}\exp(-\frac{{mv^{2}}}{2k_{B}T})$,
where $m$ is the atomic mass, $T$ is the temperature and $k_{B}$
is the Boltzmann's constant, to calculate the number of atoms with
velocity $v$. To perform the average over the thermal ensemble, we
introduce velocity-dependent detunings: $\tilde{\Delta}_{2}^{(v)}=\tilde{\Delta}_{2}+\frac{{2\pi v}}{\lambda_{1}}$,
$\tilde{\Delta}_{3}^{(v)}=\tilde{\Delta}_{3}+\frac{{2\pi v}}{\lambda_{1}}+\frac{{2\pi v}}{\lambda_{2}}$,
$\tilde{\Delta}_{4}^{(v)}=\tilde{\Delta}_{4}+\frac{{2\pi v}}{\lambda_{4}}$. Then
we integrate the optical polarization $P_{3}$ with the velocity distribution
$g(v)$, obtaining:

\begin{equation}
\langle P_{3}\rangle_{T}=\int_{-\infty}^{+\infty}\sum_{\mathcal{P}}g(v)n(\mathcal{P})\mu_{43}^{*}(\mathcal{P})\rho_{43}(\mathcal{P},v)\mathrm{d}v+\mathrm{c.c.},\label{eq:P3=00003Dintdv_sumpaths}
\end{equation}
where we have neglected velocity-dependent pumping effects and assumed
the distribution of the atoms among ground levels $n(\mathcal{P})$
to be velocity independent.

In the perturbative case $\Omega\ll\Gamma$, we can perform
the above integration analytically. Since only the susceptibility
$\chi(\mathcal{P})$ given in Eq.~(\ref{eq:chi=00003Dmu/Delta})
is velocity dependent, the velocity-averaged version of Eq.~(\ref{eq:P3=00003DEEEnchi})
is $\langle P_{3}\rangle_{T}=\epsilon_0 E_{1}E_{2}E_{4}^{*}\sum_{\mathcal{P}}\langle\chi(\mathcal{P})\rangle_{T}+\mathrm{c.c.}$
Here the susceptibility $\langle\chi(\mathcal{P})\rangle_{T}$ is
calculated by integrating $g(v)/\tilde{\Delta}_{2}^{(v)}\tilde{\Delta}_{3}^{(v)}\tilde{\Delta}_{4}^{(v)*}$, which in turn is reduced into a sum of Voigt-type profile integrals.
This is achieved by partial fractions decomposition as detailed in
the Appendix. The result can be cast into the following form: 
\begin{multline}
\langle\chi(\mathcal{P})\rangle_{T}=n(\mathcal{P})\frac{\mu_{12}(\mathcal{P})\mu_{23}(\mathcal{P})\mu_{43}^*(\mathcal{P})\mu_{14}^*(\mathcal{P})}{16 i \epsilon_0 \hbar^3}\times \\ \sqrt{{\frac{{m}}{2\pi k_{B}T}}} \sum_{i}\frac{\Lambda_{i}\mathcal{V}\left(\tilde{\Delta}_{i}(\mathcal{P})\frac{\Lambda_{i}}{2\pi}\sqrt{{\frac{{k_{B}T}}{2m}}}\right)}{Q_{i}(\{\tilde{\Delta}_{j}(\mathcal{P})\})},\label{eq:Avgchi=00003Dw/Q}
\end{multline}
where $\mathcal{V}(z)$ is the profile function we define in the Appendix, $\Lambda_2=\lambda_1$, $\Lambda_3=\frac{\lambda_1 \lambda_2}{\lambda_1+\lambda_2}$, $\Lambda_4=\lambda_4$ and $Q_{i}(\{\tilde{\Delta}_{j}\})$ are second order polynomials
of detunings. The polynomials $Q_{i}(\{\tilde{\Delta}_{j}\})$, given
in the Appendix as well, are responsible for the sub-Doppler features
in the 4WM spectrum. Their real parts are zero when real parts of
any two out of three factors in the susceptibility denominator $\tilde{\Delta}_{2}^{(v)}\tilde{\Delta}_{3}^{(v)}\tilde{\Delta}_{4}^{(v)*}$
are zero for the same velocity $v$. This determines the position of the
resonances, as we will see in Sec. \ref{sec:Results}.

\section{Experiment\label{sec:Experiment}}

\begin{figure}
\includegraphics{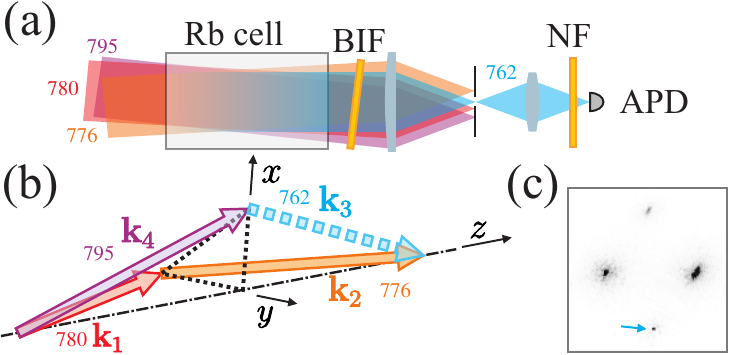}

\protect\caption{(Color online) Schematic of the experimental situation. Generation and detection
stage (a) of the experimental setup; beams enter the rubidium vapor
cell and the four-wave mixing signal is separated from the driving light
using spatial separation, band-pass interference (BIF) filter and
a notch filter (NF). The signal is detected using an avalanche photodiode
(APD). The 3D configuration of beams (b) enables phase matching without
beam overlap. In the focal plane (c) we see four distinct points corresponding
to four beams. The signal at 762~nm is marked with an arrow.}

\label{fig:exper}
\end{figure}

The heart of the experimental setup is a magnetically shielded rubidium
vapor cell heated to the temperature of 373~K. The 2.5-cm-diameter
and 7.5-cm-long cell contains a natural mixture of rubidium isotopes
and no buffer gas. We have decided to use only $^{85}\mathrm{Rb}$
(nuclear spin $\mathcal{I}=5/2$) due to its higher concentration
and, in consequence, higher optical depth. For the ground state $|1\rangle$
we have chosen the $F_{1}=3$ hyperfine component of the $5^{2}S_{1/2}$
state. For the intermediate states we have chosen $5^{2}P_{3/2}$
for the $|2\rangle$ state and $5^{2}P_{1/2}$ for the last intermediate
state $|4\rangle$. As the highest excited state $|3\rangle$ we use the
$5^{2}D_{3/2}$ level. The respective wavelengths for the resonant
transitions are $\lambda_{1}$=780~nm (the D2 line), $\lambda_{2}$=776~nm, $\lambda_{3}$=762~nm and $\lambda_{4}$=795~nm (the D1 line).
Out of many possible diamond configurations the one we use has the
main advantage of high efficiency of detectors for each of the wavelengths,
thus enabling future quantum optics applications.

Inside the cell three beams from three different lasers at 780, 795,
and 776~nm intersect at a small angle as depicted in Fig.~\ref{fig:exper}.
The fourth beam of light at 762~nm is generated in the cell according
to the phase matching condition $\mathbf{{k_{1}}}+\mathbf{{k_{2}}}=\mathbf{{k_{3}}}+\mathbf{{k_{4}}}$.
The values of wave vectors are approximately $\mathbf{k_{1}}\approx\frac{2\pi}{\lambda_{1}}[0,-\theta,1]$,
$\mathbf{k_{2}}\approx\frac{2\pi}{\lambda_{2}}[0,\theta,1]$, $\mathbf{{k_{3}}}\approx\frac{2\pi}{\lambda_{3}}[-\theta,0,1]$,
and $\mathbf{{k_{4}}}\approx\frac{2\pi}{\lambda_{4}}[\theta,0,1]$,
where $\theta=8$~mrad. The nearly collinear configuration entails broad phase matching. Experimentally, we verified that varying the angle of $\mathbf{k_1}$ by 2~mrad shifts the signal beam accordingly but does not have significant influence on signal intensity. Inside the cell, the beams are collimated with a $1/e^{2}$ diameter of approximately 3~mm.

We use three Toptica lasers: two distributed feedback diodes
at 780 and 795~nm and an external cavity laser at 776~nm. The beams
are combined to intersect at a small angle inside the cell and then
we employ a high-extinction Wollaston polarizer to prepare the light
of all beams in the $x$ polarization state. 

Finally, the generated 762~nm light needs to be separated from the
driving beams. We use three distinct filtering methods to obtain high
signal-to-background ratio, as shown in Fig.~\ref{fig:exper}(c).
A tilted interference bandpass filter comes first, then we use an
iris diaphragm in the focal plane to cover the driving beams, and finally
we apply a notch filter with a central wavelength of 785~nm (Thorlabs
NF785-33). The only residual light from the driving lasers we detect
comes from the amplified spontaneous emission at 762~nm in laser diodes.
The 4WM signal is detected using an avalanche photodiode (Thorlabs
APD120A). 

The measurement scheme is designed to obtain maps of the 4WM signal
as a function of two out of three driving laser detunings. We lock the
frequency of the 776-nm laser using a wavelength meter (Angstrom WS-7)
at a fixed detuning $\Delta_{776}$, nearly resonant to the $|2\rangle-|3\rangle$
transition. During the measurement, the detuning $\Delta_{795}$ of
the 795-nm laser frequency is altered in small steps of $4\times2\pi$~MHz, while the detuning $\Delta_{780}$ of the 780-nm laser is scanned
over the relevant range of optical frequencies. The exact scan rate
varied for different measurements, but was of the order of $50\times2\pi$~MHz/ms. Data from 800 to 1000 scans were collected and averaged to
give a dependence of the 4WM signal from $\Delta_{780}$ for fixed
$\Delta_{795}$ and $\Delta_{776}$. 

All detunings are measured from the centroid of respective resonance
line. The detunings of lasers resonant to D1 and D2 lines are determined
using saturated absorption signal obtained in auxiliary rubidium vapor
cells. Note that through comparing the laser detunings with the theoretical
model, we obtain $\Delta_{2}=\Delta_{780}$, $\Delta_{3}=\Delta_{780}+\Delta_{776}$
and $\Delta_{4}=\Delta_{795}$.

\section{Results\label{sec:Results}}

\begin{figure}
\includegraphics[scale=0.7]{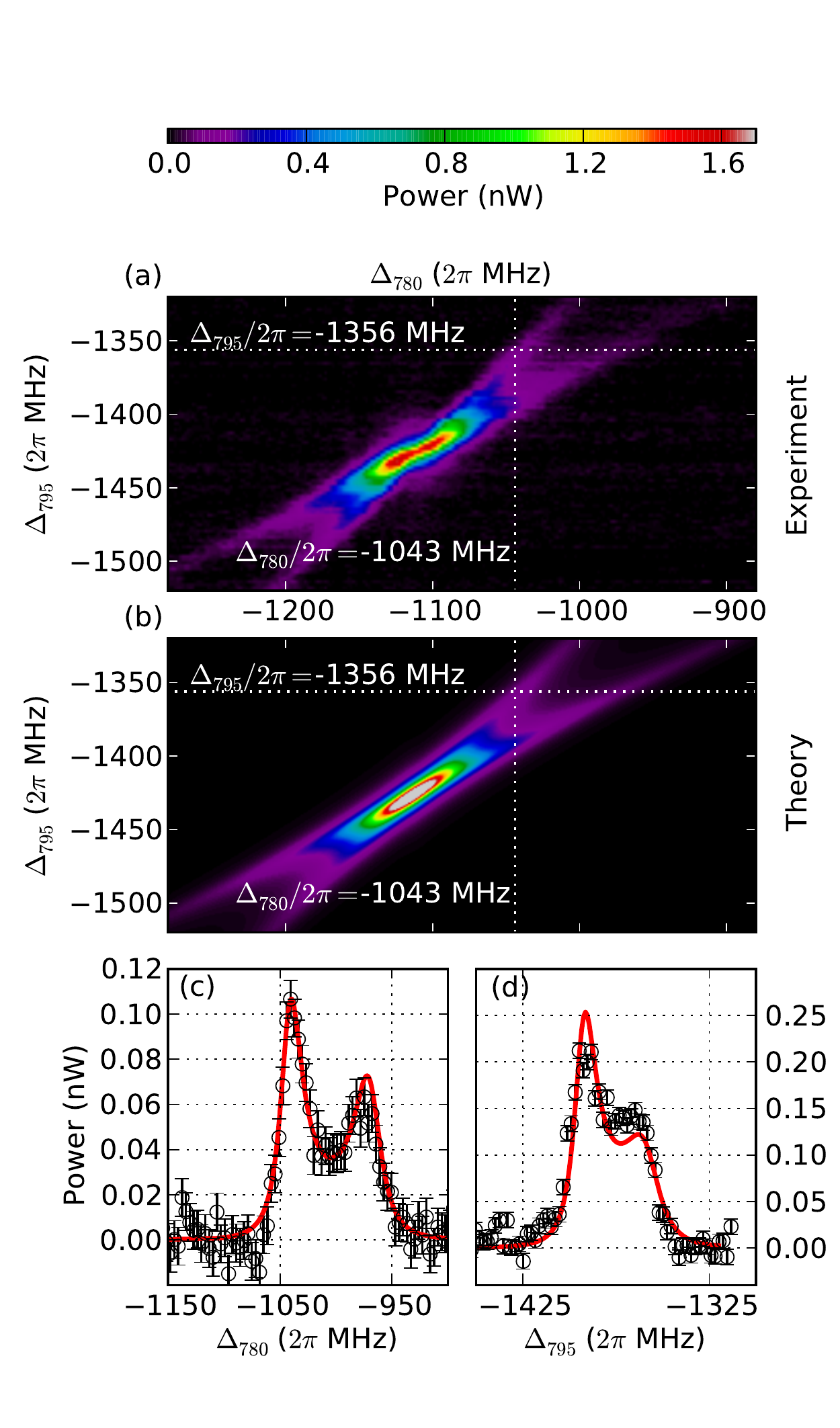}

\raggedright{}\protect\caption{(Color online) Experimental (a) and theoretical (b) maps of the intensity of the
4WM signal $|E_3|^2$ as a function of 780~nm and 795~nm laser detunings in case
of low driving field intensities around the $F_{2}=2$ and $F_{4}=4$ resonance.
The cross-sections, marked on the maps with dashed lines, from experimental (dots with errorbars) and theoretical
(solid lines) data are presented in figures (c) and (d), where (c) depicts the horizontal and (d) the vertical cross-section. The detuning
of 776~nm laser frequency was $\Delta_{776}/2\pi=-51$~MHz.\label{fig:et-p}}
\end{figure}

The first measurement we present was aimed to determine the shape
of a single 4WM resonance in the regime of a perturbative solution.
To verify the linear perturbative theory we performed a measurement
of the 4WM signal, where all driving field intensities were low and corresponded
to the Rabi frequencies of $\Omega_{1}/2\pi$=5~MHz, $\Omega_{2}/2\pi$=0.3
MHz and $\Omega_{4}/2\pi$=4~MHz. Note that these are all lower than
corresponding transitions' linewidths. We chose the strongest resonance
that corresponded to the path leading through $F_{2}=4$ , $F_{3}=3$,
and then $F_{4}=2$ levels. The choice of such a configuration entails
that there should be no splitting of this resonance due to the hyperfine
structure of the highest $|3\rangle$ state, as there is only one
possible spin $F_{3}$ if $F_{2}=4$ and $F_{4}=2$. 

Figures \ref{fig:et-p}(a) and \ref{fig:et-p}(b) present the experimental
and theoretical 4WM signal intensity $|E_3|^2$, respectively. The presented
maps show an excellent conformity of theory [Eq.~(\ref{eq:Avgchi=00003Dw/Q})]
and experiment. The only free parameter for the theory is the intensity
multiplicative factor. Additionally, we need to take into account
the laser linewidth, which is of the order of several~MHz for each
laser when averaged over duration of a single measurement.

In Figs. \ref{fig:et-p}(c) and \ref{fig:et-p}(d), we present two
cross sections of the maps. The cross-sections, denoted by dashed
lines on the maps, are taken far from the resonance, as the structure
there is non-trivial. It exhibits two peaks and it is instructive
to observe their separation and relative intensities. The cross sections
show a quantitative agreement between theory (solid line) and experiment
(dots with error bars), confirming the correctness of our theoretical
approach. On the maps we observe a small discrepancy of intensity and
shape in the very center of the resonance that are due to small non-linearity,
as the Rabi frequencies are only slightly lower than the linewidths.

The results of the second measurement we present here demonstrate the
influence of the hyperfine structure of intermediate levels on the
4WM signal. For this measurement we increased the power of each of
the driving lasers, and obtained Rabi frequencies of $\Omega_{1}/2\pi$=34
MHz, $\Omega_{2}/2\pi$=0.7~MHz and $\Omega_{4}/2\pi$=28~MHz. As
the ground state is the $F_{1}=3$ state, we expect six possible resonances,
as this ground state is coupled to three hyperfine levels ($F_{2}=2,3,4$)
of the $|2\rangle$ state and two hyperfine levels ($F_{4}=2,3$)
of the $|4\rangle$ state. We do not expect to see different resonances
arising due to the hyperfine structure of the highest state $|3\rangle$,
as the hyperfine splitting of this state is smaller than the resolution
of our experiment. 

\begin{figure}
\includegraphics[scale=0.7]{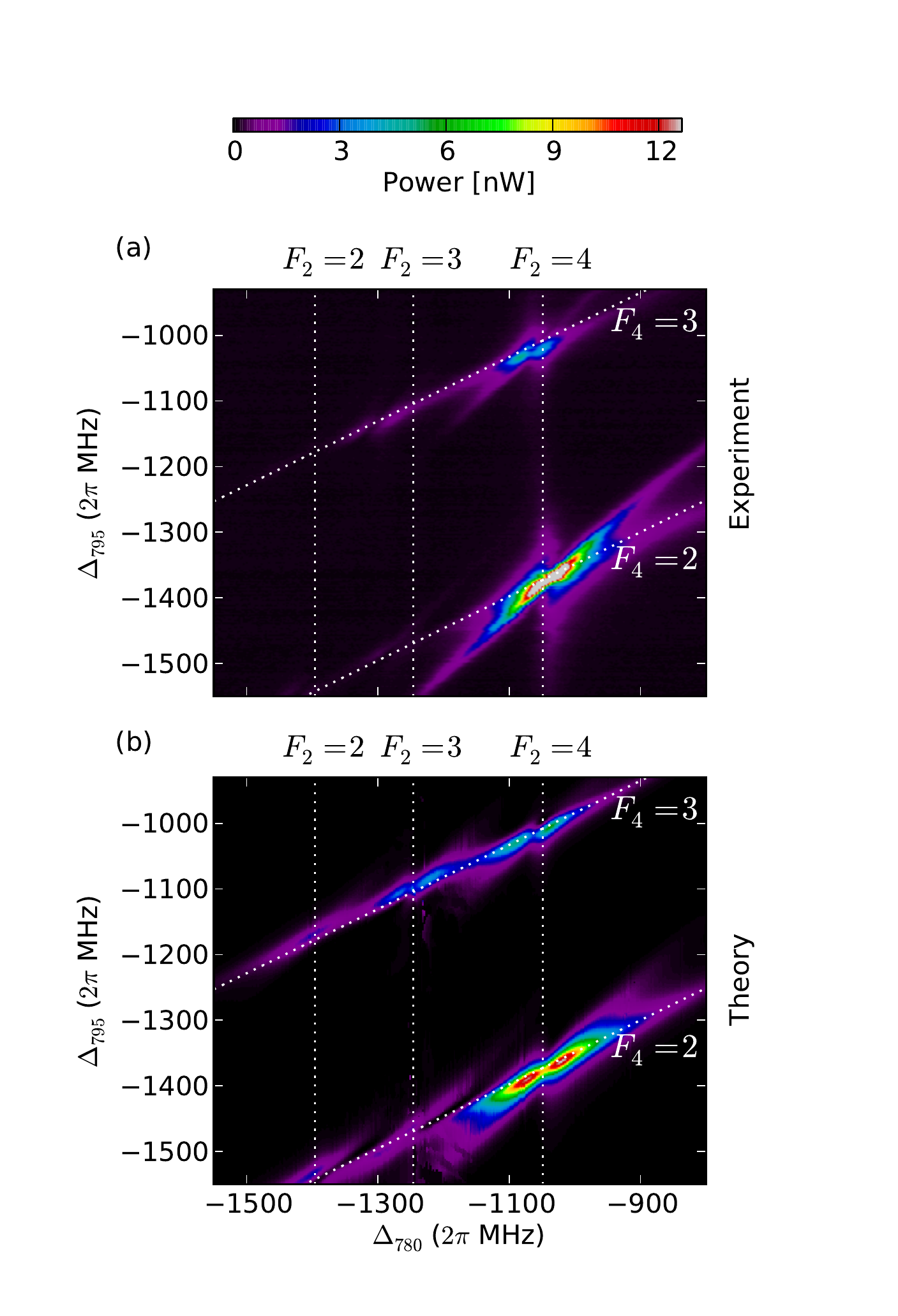}\protect\caption{(Color online) Experimental (a) and theoretical (b) 4WM signal demonstrating all
possible resonances arising due to the hyperfine structure. Dashed lines correspond to the resonance conditions given by equations (\ref{eq:rescon1}) and (\ref{eq:rescon2}) for different $F_2$ and $F_4$. Crossings of the lines correspond the four-wave mixing resonances. The detuning
of 776~nm laser frequency was $\Delta_{776}/2\pi=30$~MHz. \label{fig:et-e}}
\end{figure}

Figures \ref{fig:et-e}(a) and \ref{fig:et-e}(b) present the experimental
and theoretical results for the 4WM signal, respectively. We identify
the positions of the resonances corresponding to the crossings of
the dashed lines. Each line corresponds to a certain hyperfine level
of the $|2\rangle$ or the $|4\rangle$ state; these lines correspond
to zeros of the real part of denominator of Eq.~(\ref{eq:Avgchi=00003Dw/Q}),
or more precisely the polynomials $L_{ij}$, defined in the Appendix.
This brings about two resonance conditions:

\begin{equation}
\Delta_{780}=-\Delta F_{1}+\frac{{\lambda_{1}+\lambda_{2}}}{\lambda_{1}}\Delta F_{2}+\frac{{\lambda_{1}}}{\lambda_{2}}\Delta_{776}\label{eq:rescon1},
\end{equation}

\begin{multline}
\Delta_{795}=\frac{{\lambda_{1}\lambda_{2}}}{(\lambda_{1}+\lambda_{2})\lambda_{4}}(\Delta_{780}+\Delta_{776})+\\ \left(\frac{{\lambda_{1}\lambda_{2}}}{(\lambda_{1}+\lambda_{2})\lambda_{4}}-1\right)\Delta F_{1}+\Delta F_{4},\label{eq:rescon2}
\end{multline}
where again we neglected hyperfine splitting of the highest excited
state $|3\rangle$. The first equation defines the lines corresponding
to hyperfine levels of the $|2\rangle$ state. These are vertical
on our map, but their separation is twice the hyperfine splitting,
as $\frac{{\lambda_{1}+\lambda_{2}}}{\lambda_{1}}\approx2$. The second
equation defines the lines corresponding to the $|4\rangle$ state.
These lines have a slope of $\frac{{\lambda_{1}\lambda_{2}}}{(\lambda_{1}+\lambda_{2})\lambda_{4}}\approx\frac{{1}}{2}$,
but their separation is simply equal to the splitting of the $|4\rangle$
state.

For the full theoretical calculation [Fig.~\ref{fig:et-e}(b)], we
use the non-perturbative framework introduced in Sec. \ref{sec:Theory}.
The results show good qualitative agreement with the experiment. We
observe that both in theory and experiment, the strongest resonances
are the rightmost ones on our map, corresponding to highest $F_{2}$
spin. This is due to the relative amplitudes of resonances, that can
be calculated even in the perturbative regime from the products of
dipole moments. In Table~\ref{tab:tab}, we give these relative amplitudes
for each resonance as a function of the ground state spin projection
$|m_{F_{1}}|.$ Note, that different ground-state populations give
rise to different 4WM resonances. Consequently, non-equilibrium populations
may significantly change observed relative amplitudes. In our calculation,
we take this into account by modifying ground state populations according
to the numerical solution of Liouville equation with relaxation and
repopulation \cite{Happer2010}.

One of the nonlinear effects we predict theoretically and observe
experimentally is Autler-Townes-type splitting \cite{Becerra2008,Autler1955}
of the $F_{2}=4$, $F_{4}=3$ resonance, that does not occur in the
linear regime. Another effect observed near the $F_{2}=4$, $F_{4}=2$
resonance is bending of the resonance line towards the vertical direction.
Namely, we note relatively higher intensity as compared to the linear
case in the top-left and bottom-right corners of the resonance.

In the experiment, we witness two more effects that are not present
in the theoretical results. The first effect is that the leftmost
resonances, corresponding to lowest $F_{2}$ spin values, are even
less intense than predicted. The second effect is the appearance of
narrow diagonal lines in the 4WM signal, which should be less intense
according to our theory. We believe that these two effects stem from
velocity-selective optical pumping, that our model does not take into
account at all. Nevertheless, we find the agreement between our theoretical
and experimental results satisfactory.

\section{Conclusions\label{sec:Conclusions}}

Beginning with the theoretical description of a four-level atom, we
have derived a model for the the intensity of 4WM in a multi-level
atomic medium with inhomogeneous broadening. We are able to take into account the interference of many possible paths of 4WM with a simple formalism, that could be easily extended to many similar situations, including possibly higher-order transitions. Our model is a reasonable
compromise between simplicity and accuracy for predicting intensities,
shapes, and positions of 4WM resonances.

In the experimental part, we have demonstrated the influence of the
rich level structure of rubidium on the 4WM process. We have obtained
excellent agreement between theory and experiment for low driving field
intensities, as well as good qualitative agreement for higher driving field
intensities, where more non-linear processes contribute.

Combined with optical pumping, our methods may facilitate the design
of coherent control in a multi-level atomic medium with light. We
have shown that light at 762 nm couples to the population of the
ground state. Future investigations could include coupling of 762-nm light to the ground state coherence in order to create Raman-scattering-like
light-atom interface \cite{Chrapkiewicz2012} or engineering the 4WM
signal by optically pumping the ground-state. This could be accomplished
in the cells with buffer gas that makes the atomic motion diffusive
\cite{Parniak2013a}, the cell with anti-relaxation coating \cite{Balabas2010},
or in a cold-atomic ensemble. Another approach could involve control of 4WM signal through velocity-selective optical pumping.

\section*{Acknowledgments}

We acknowledge the generous support from Konrad Banaszek, Czes\l aw
Radzewicz and Tadeusz Stacewicz, as well as insightful discussions with Rados\l aw Chrapkiewicz. This work was supported by the National
Science Center Grant No. DEC-2011/03/D/ST2/01941 and by the Polish
Ministry of Science and Higher Education ``Diamentowy Grant'' Project
No. DI2013 011943.

\section*{Appendix: Ensemble-averaged susceptibility\label{Appendix}}

We decompose the formula~(\ref{eq:coherence-p}) for the optical
coherence $\rho_{43}$ in the perturbative regime with velocity-dependent
detunings, obtaining a sum of three expressions with first order polynomials
in terms of velocity in the denominators:
\begin{equation}
\left(\tilde{\Delta}_{2}^{(v)}\tilde{\Delta}_{3}^{(v)}\tilde{\Delta}_{4}^{(v)*}\right)^{-1}=\sum_{i}\left(\tilde{\Delta}_{i}^{(v)}Q_{i}(\{\tilde{\Delta}_{j}\})\right)^{-1}
\end{equation}
\onecolumngrid

The polynomials are: $Q_{2}=L_{23}L_{24}$, $Q_{3}=\frac{{\lambda_{4}(\lambda_{1}+\lambda_{2})^{2}}}{\lambda_{1}\lambda_{2}^{2}}L_{23}L_{34}$,
$Q_{4}=\frac{{\lambda_{1}}}{\lambda_{2}}L_{24}L_{34}$, where $L_{23}=(1+\frac{{\lambda_{1}}}{\lambda_{2}})\tilde{\Delta}_{2}-\tilde{\Delta}_{3}$,
$L_{24}=\frac{{\lambda_{1}}}{\lambda_{2}}\tilde{\Delta}_{2}-\tilde{\Delta_{4}}$
and $L_{34}=\lambda_{4}\left(\frac{{1}}{\lambda_{1}}+\frac{{1}}{\lambda_{2}}\right)\tilde{\Delta}_{4}-\tilde{\Delta}_{3}$.
Equating the real part of $L_{ij}$ to zero yields resonance conditions
given by Eq.~(\ref{eq:rescon1}) and~(\ref{eq:rescon2}). The expression
can now be easily integrated with velocity distribution $g(v)$, as
each component can be integrated separately to give the Voigt-like profile
used in Eq.~(\ref{eq:Avgchi=00003Dw/Q}):

\begin{equation}
\int_{-\infty}^{\infty}\frac{e^{-v^{2}/2\langle v^{2}\rangle}}{2\pi v/\Lambda+\tilde{\Delta}}\mathrm{{d}}v=\frac{\Lambda}{2i}\exp\left(-\frac{{\tilde{\Delta}^{2}}\left(\frac{\Lambda}{2\pi}\right)^{2}}{2\langle v^{2}\rangle}\right)\left(\mathrm{{erf}}\left(i\frac{{\tilde{\Delta}\frac{\Lambda}{2\pi}}}{\sqrt{2\langle v^{2}\rangle}}\right)\pm1\right)=\frac{\Lambda}{2i}\mathcal{V}\left(\frac{{\tilde{{\Delta}}\frac{\Lambda}{2\pi}}}{\sqrt{{2\langle v^{2}\rangle}}}\right),
\end{equation}

where we take $+1$ for $\mathrm{Im}(\tilde{{\Delta}})>0$ and $-1$ otherwise
and $\langle v^{2}\rangle=\frac{{k_{B}T}}{m}$. The profile function
is defined as:
\begin{equation}
\mathcal{{V}}(z)=\begin{cases}
e^{-z^{2}}(\mathrm{{erf}}(iz)+1) & \mathrm{Im}(z)>0\\
e^{-z^{2}}(\mathrm{{erf}}(iz)-1) & \mathrm{Im}(z)<0
\end{cases}
\end{equation}
Note that it is customary to use the Faddeeva function $w(z)=e^{-z^{2}}(\mathrm{{erf}(i}z)+1)$
\cite{Happer2010}, but neither this function nor the plasma dispersion
function can be used here, as they would yield incorrect results in the lower
half of the complex plane.
\twocolumngrid
\bibliographystyle{apsrev4-1}
\bibliography{bibliography}

\begin{thebibliography}{30}%
\makeatletter
\providecommand \@ifxundefined [1]{%
 \@ifx{#1\undefined}
}%
\providecommand \@ifnum [1]{%
 \ifnum #1\expandafter \@firstoftwo
 \else \expandafter \@secondoftwo
 \fi
}%
\providecommand \@ifx [1]{%
 \ifx #1\expandafter \@firstoftwo
 \else \expandafter \@secondoftwo
 \fi
}%
\providecommand \natexlab [1]{#1}%
\providecommand \enquote  [1]{``#1''}%
\providecommand \bibnamefont  [1]{#1}%
\providecommand \bibfnamefont [1]{#1}%
\providecommand \citenamefont [1]{#1}%
\providecommand \href@noop [0]{\@secondoftwo}%
\providecommand \href [0]{\begingroup \@sanitize@url \@href}%
\providecommand \@href[1]{\@@startlink{#1}\@@href}%
\providecommand \@@href[1]{\endgroup#1\@@endlink}%
\providecommand \@sanitize@url [0]{\catcode `\\12\catcode `\$12\catcode
  `\&12\catcode `\#12\catcode `\^12\catcode `\_12\catcode `\%12\relax}%
\providecommand \@@startlink[1]{}%
\providecommand \@@endlink[0]{}%
\providecommand \url  [0]{\begingroup\@sanitize@url \@url }%
\providecommand \@url [1]{\endgroup\@href {#1}{\urlprefix }}%
\providecommand \urlprefix  [0]{URL }%
\providecommand \Eprint [0]{\href }%
\providecommand \doibase [0]{http://dx.doi.org/}%
\providecommand \selectlanguage [0]{\@gobble}%
\providecommand \bibinfo  [0]{\@secondoftwo}%
\providecommand \bibfield  [0]{\@secondoftwo}%
\providecommand \translation [1]{[#1]}%
\providecommand \BibitemOpen [0]{}%
\providecommand \bibitemStop [0]{}%
\providecommand \bibitemNoStop [0]{.\EOS\space}%
\providecommand \EOS [0]{\spacefactor3000\relax}%
\providecommand \BibitemShut  [1]{\csname bibitem#1\endcsname}%
\let\auto@bib@innerbib\@empty
\bibitem [{\citenamefont {Peyronel}\ \emph {et~al.}(2012)\citenamefont
  {Peyronel}, \citenamefont {Firstenberg}, \citenamefont {Liang}, \citenamefont
  {Hofferberth}, \citenamefont {Gorshkov}, \citenamefont {Pohl}, \citenamefont
  {Lukin},\ and\ \citenamefont {Vuleti\'{c}}}]{Peyronel2012}%
  \BibitemOpen
  \bibfield  {author} {\bibinfo {author} {\bibfnamefont {T.}~\bibnamefont
  {Peyronel}}, \bibinfo {author} {\bibfnamefont {O.}~\bibnamefont
  {Firstenberg}}, \bibinfo {author} {\bibfnamefont {Q.-Y.}\ \bibnamefont
  {Liang}}, \bibinfo {author} {\bibfnamefont {S.}~\bibnamefont {Hofferberth}},
  \bibinfo {author} {\bibfnamefont {A.~V.}\ \bibnamefont {Gorshkov}}, \bibinfo
  {author} {\bibfnamefont {T.}~\bibnamefont {Pohl}}, \bibinfo {author}
  {\bibfnamefont {M.~D.}\ \bibnamefont {Lukin}}, \ and\ \bibinfo {author}
  {\bibfnamefont {V.}~\bibnamefont {Vuleti\'{c}}},\ }\href {\doibase
  10.1038/nature11361} {\bibfield  {journal} {\bibinfo  {journal} {Nature}\
  }\textbf {\bibinfo {volume} {488}},\ \bibinfo {pages} {57} (\bibinfo {year}
  {2012})}\BibitemShut {NoStop}%
\bibitem [{\citenamefont {Clow}\ and\ \citenamefont
  {Weinacht}(2010)}]{Clow2010}%
  \BibitemOpen
  \bibfield  {author} {\bibinfo {author} {\bibfnamefont {S.}~\bibnamefont
  {Clow}}\ and\ \bibinfo {author} {\bibfnamefont {T.}~\bibnamefont
  {Weinacht}},\ }\href {\doibase 10.1103/PhysRevA.82.023411} {\bibfield
  {journal} {\bibinfo  {journal} {Phys. Rev. A}\ }\textbf {\bibinfo {volume}
  {82}},\ \bibinfo {pages} {023411} (\bibinfo {year} {2010})}\BibitemShut
  {NoStop}%
\bibitem [{\citenamefont {Lee}\ \emph {et~al.}(2013)\citenamefont {Lee},
  \citenamefont {Kim}, \citenamefont {Lim},\ and\ \citenamefont
  {Ahn}}]{Lee2013}%
  \BibitemOpen
  \bibfield  {author} {\bibinfo {author} {\bibfnamefont {H.-g.}\ \bibnamefont
  {Lee}}, \bibinfo {author} {\bibfnamefont {H.}~\bibnamefont {Kim}}, \bibinfo
  {author} {\bibfnamefont {J.}~\bibnamefont {Lim}}, \ and\ \bibinfo {author}
  {\bibfnamefont {J.}~\bibnamefont {Ahn}},\ }\href {\doibase
  10.1103/PhysRevA.88.053427} {\bibfield  {journal} {\bibinfo  {journal} {Phys.
  Rev. A}\ }\textbf {\bibinfo {volume} {88}},\ \bibinfo {pages} {053427}
  (\bibinfo {year} {2013})}\BibitemShut {NoStop}%
\bibitem [{\citenamefont {Sell}\ \emph {et~al.}(2014)\citenamefont {Sell},
  \citenamefont {Gearba}, \citenamefont {DePaola},\ and\ \citenamefont
  {Knize}}]{Sell2014}%
  \BibitemOpen
  \bibfield  {author} {\bibinfo {author} {\bibfnamefont {J.~F.}\ \bibnamefont
  {Sell}}, \bibinfo {author} {\bibfnamefont {M.~A.}\ \bibnamefont {Gearba}},
  \bibinfo {author} {\bibfnamefont {B.~D.}\ \bibnamefont {DePaola}}, \ and\
  \bibinfo {author} {\bibfnamefont {R.~J.}\ \bibnamefont {Knize}},\ }\href
  {\doibase 10.1364/OL.39.000528} {\bibfield  {journal} {\bibinfo  {journal}
  {Opt. Lett.}\ }\textbf {\bibinfo {volume} {39}},\ \bibinfo {pages} {528}
  (\bibinfo {year} {2014})}\BibitemShut {NoStop}%
\bibitem [{\citenamefont {Meijer}\ \emph {et~al.}(2006)\citenamefont {Meijer},
  \citenamefont {White}, \citenamefont {Smeets}, \citenamefont {Jeppesen},\
  and\ \citenamefont {Scholten}}]{Meijer2006}%
  \BibitemOpen
  \bibfield  {author} {\bibinfo {author} {\bibfnamefont {T.}~\bibnamefont
  {Meijer}}, \bibinfo {author} {\bibfnamefont {J.~D.}\ \bibnamefont {White}},
  \bibinfo {author} {\bibfnamefont {B.}~\bibnamefont {Smeets}}, \bibinfo
  {author} {\bibfnamefont {M.}~\bibnamefont {Jeppesen}}, \ and\ \bibinfo
  {author} {\bibfnamefont {R.~E.}\ \bibnamefont {Scholten}},\ }\href {\doibase
  10.1364/OL.31.001002} {\bibfield  {journal} {\bibinfo  {journal} {Opt.
  Lett.}\ }\textbf {\bibinfo {volume} {31}},\ \bibinfo {pages} {1002} (\bibinfo
  {year} {2006})}\BibitemShut {NoStop}%
\bibitem [{\citenamefont {Zibrov}\ \emph {et~al.}(2002)\citenamefont {Zibrov},
  \citenamefont {Lukin}, \citenamefont {Hollberg},\ and\ \citenamefont
  {Scully}}]{Zibrov2002}%
  \BibitemOpen
  \bibfield  {author} {\bibinfo {author} {\bibfnamefont {A.~S.}\ \bibnamefont
  {Zibrov}}, \bibinfo {author} {\bibfnamefont {M.~D.}\ \bibnamefont {Lukin}},
  \bibinfo {author} {\bibfnamefont {L.}~\bibnamefont {Hollberg}}, \ and\
  \bibinfo {author} {\bibfnamefont {M.~O.}\ \bibnamefont {Scully}},\ }\href
  {\doibase 10.1103/PhysRevA.65.051801} {\bibfield  {journal} {\bibinfo
  {journal} {Phys. Rev. A}\ }\textbf {\bibinfo {volume} {65}},\ \bibinfo
  {pages} {051801} (\bibinfo {year} {2002})}\BibitemShut {NoStop}%
\bibitem [{\citenamefont {Akulshin}\ \emph {et~al.}(2009)\citenamefont
  {Akulshin}, \citenamefont {McLean}, \citenamefont {Sidorov},\ and\
  \citenamefont {Hannaford}}]{Akulshin2009a}%
  \BibitemOpen
  \bibfield  {author} {\bibinfo {author} {\bibfnamefont {A.~M.}\ \bibnamefont
  {Akulshin}}, \bibinfo {author} {\bibfnamefont {R.~J.}\ \bibnamefont
  {McLean}}, \bibinfo {author} {\bibfnamefont {A.~I.}\ \bibnamefont {Sidorov}},
  \ and\ \bibinfo {author} {\bibfnamefont {P.}~\bibnamefont {Hannaford}},\
  }\href {\doibase 10.1364/OE.17.022861} {\bibfield  {journal} {\bibinfo
  {journal} {Opt. Express}\ }\textbf {\bibinfo {volume} {17}},\ \bibinfo
  {pages} {22861} (\bibinfo {year} {2009})}\BibitemShut {NoStop}%
\bibitem [{\citenamefont {Morigi}\ \emph {et~al.}(2002)\citenamefont {Morigi},
  \citenamefont {Franke-Arnold},\ and\ \citenamefont {Oppo}}]{Morigi2002}%
  \BibitemOpen
  \bibfield  {author} {\bibinfo {author} {\bibfnamefont {G.}~\bibnamefont
  {Morigi}}, \bibinfo {author} {\bibfnamefont {S.}~\bibnamefont
  {Franke-Arnold}}, \ and\ \bibinfo {author} {\bibfnamefont {G.-L.}\
  \bibnamefont {Oppo}},\ }\href {\doibase 10.1103/PhysRevA.66.053409}
  {\bibfield  {journal} {\bibinfo  {journal} {Phys. Rev. A}\ }\textbf {\bibinfo
  {volume} {66}},\ \bibinfo {pages} {053409} (\bibinfo {year}
  {2002})}\BibitemShut {NoStop}%
\bibitem [{\citenamefont {Walker}\ \emph {et~al.}(2012)\citenamefont {Walker},
  \citenamefont {Arnold},\ and\ \citenamefont {Franke-Arnold}}]{Walker2012}%
  \BibitemOpen
  \bibfield  {author} {\bibinfo {author} {\bibfnamefont {G.}~\bibnamefont
  {Walker}}, \bibinfo {author} {\bibfnamefont {A.~S.}\ \bibnamefont {Arnold}},
  \ and\ \bibinfo {author} {\bibfnamefont {S.}~\bibnamefont {Franke-Arnold}},\
  }\href {\doibase 10.1103/PhysRevLett.108.243601} {\bibfield  {journal}
  {\bibinfo  {journal} {Phys. Rev. Lett.}\ }\textbf {\bibinfo {volume} {108}},\
  \bibinfo {pages} {243601} (\bibinfo {year} {2012})}\BibitemShut {NoStop}%
\bibitem [{\citenamefont {Chaneli\`{e}re}\ \emph {et~al.}(2006)\citenamefont
  {Chaneli\`{e}re}, \citenamefont {Matsukevich}, \citenamefont {Jenkins},
  \citenamefont {Kennedy}, \citenamefont {Chapman},\ and\ \citenamefont
  {Kuzmich}}]{Chaneliere2006}%
  \BibitemOpen
  \bibfield  {author} {\bibinfo {author} {\bibfnamefont {T.}~\bibnamefont
  {Chaneli\`{e}re}}, \bibinfo {author} {\bibfnamefont {D.~N.}\ \bibnamefont
  {Matsukevich}}, \bibinfo {author} {\bibfnamefont {S.~D.}\ \bibnamefont
  {Jenkins}}, \bibinfo {author} {\bibfnamefont {T.~A.~B.}\ \bibnamefont
  {Kennedy}}, \bibinfo {author} {\bibfnamefont {M.~S.}\ \bibnamefont
  {Chapman}}, \ and\ \bibinfo {author} {\bibfnamefont {A.}~\bibnamefont
  {Kuzmich}},\ }\href {\doibase 10.1103/PhysRevLett.96.093604} {\bibfield
  {journal} {\bibinfo  {journal} {Phys. Rev. Lett.}\ }\textbf {\bibinfo
  {volume} {96}},\ \bibinfo {pages} {093604} (\bibinfo {year}
  {2006})}\BibitemShut {NoStop}%
\bibitem [{\citenamefont {Donvalkar}\ \emph {et~al.}(2014)\citenamefont
  {Donvalkar}, \citenamefont {Venkataraman}, \citenamefont {Clemmen},
  \citenamefont {Saha},\ and\ \citenamefont {Gaeta}}]{Donvalkar2014}%
  \BibitemOpen
  \bibfield  {author} {\bibinfo {author} {\bibfnamefont {P.~S.}\ \bibnamefont
  {Donvalkar}}, \bibinfo {author} {\bibfnamefont {V.}~\bibnamefont
  {Venkataraman}}, \bibinfo {author} {\bibfnamefont {S.}~\bibnamefont
  {Clemmen}}, \bibinfo {author} {\bibfnamefont {K.}~\bibnamefont {Saha}}, \
  and\ \bibinfo {author} {\bibfnamefont {A.~L.}\ \bibnamefont {Gaeta}},\ }\href
  {\doibase 10.1364/OL.39.001557} {\bibfield  {journal} {\bibinfo  {journal}
  {Opt. Lett.}\ }\textbf {\bibinfo {volume} {39}},\ \bibinfo {pages} {1557}
  (\bibinfo {year} {2014})}\BibitemShut {NoStop}%
\bibitem [{\citenamefont {Radnaev}\ \emph {et~al.}(2010)\citenamefont
  {Radnaev}, \citenamefont {Dudin}, \citenamefont {Zhao}, \citenamefont {Jen},
  \citenamefont {Jenkins}, \citenamefont {Kuzmich},\ and\ \citenamefont
  {Kennedy}}]{Radnaev2010a}%
  \BibitemOpen
  \bibfield  {author} {\bibinfo {author} {\bibfnamefont {A.~G.}\ \bibnamefont
  {Radnaev}}, \bibinfo {author} {\bibfnamefont {Y.~O.}\ \bibnamefont {Dudin}},
  \bibinfo {author} {\bibfnamefont {R.}~\bibnamefont {Zhao}}, \bibinfo {author}
  {\bibfnamefont {H.~H.}\ \bibnamefont {Jen}}, \bibinfo {author} {\bibfnamefont
  {S.~D.}\ \bibnamefont {Jenkins}}, \bibinfo {author} {\bibfnamefont
  {A.}~\bibnamefont {Kuzmich}}, \ and\ \bibinfo {author} {\bibfnamefont
  {T.~A.~B.}\ \bibnamefont {Kennedy}},\ }\href {\doibase 10.1038/nphys1773}
  {\bibfield  {journal} {\bibinfo  {journal} {Nat. Phys.}\ }\textbf {\bibinfo
  {volume} {6}},\ \bibinfo {pages} {894} (\bibinfo {year} {2010})}\BibitemShut
  {NoStop}%
\bibitem [{\citenamefont {Jen}\ and\ \citenamefont {Kennedy}(2010)}]{Jen2010}%
  \BibitemOpen
  \bibfield  {author} {\bibinfo {author} {\bibfnamefont {H.~H.}\ \bibnamefont
  {Jen}}\ and\ \bibinfo {author} {\bibfnamefont {T.~A.~B.}\ \bibnamefont
  {Kennedy}},\ }\href {\doibase 10.1103/PhysRevA.82.023815} {\bibfield
  {journal} {\bibinfo  {journal} {Phys. Rev. A}\ }\textbf {\bibinfo {volume}
  {82}},\ \bibinfo {pages} {023815} (\bibinfo {year} {2010})}\BibitemShut
  {NoStop}%
\bibitem [{\citenamefont {Willis}\ \emph {et~al.}(2009)\citenamefont {Willis},
  \citenamefont {Becerra}, \citenamefont {Orozco},\ and\ \citenamefont
  {Rolston}}]{Willis2009a}%
  \BibitemOpen
  \bibfield  {author} {\bibinfo {author} {\bibfnamefont {R.~T.}\ \bibnamefont
  {Willis}}, \bibinfo {author} {\bibfnamefont {F.~E.}\ \bibnamefont {Becerra}},
  \bibinfo {author} {\bibfnamefont {L.~A.}\ \bibnamefont {Orozco}}, \ and\
  \bibinfo {author} {\bibfnamefont {S.~L.}\ \bibnamefont {Rolston}},\ }\href
  {\doibase 10.1103/PhysRevA.79.033814} {\bibfield  {journal} {\bibinfo
  {journal} {Phys. Rev. A}\ }\textbf {\bibinfo {volume} {79}},\ \bibinfo
  {pages} {033814} (\bibinfo {year} {2009})}\BibitemShut {NoStop}%
\bibitem [{\citenamefont {Becerra}\ \emph {et~al.}(2008)\citenamefont
  {Becerra}, \citenamefont {Willis}, \citenamefont {Rolston},\ and\
  \citenamefont {Orozco}}]{Becerra2008}%
  \BibitemOpen
  \bibfield  {author} {\bibinfo {author} {\bibfnamefont {F.~E.}\ \bibnamefont
  {Becerra}}, \bibinfo {author} {\bibfnamefont {R.~T.}\ \bibnamefont {Willis}},
  \bibinfo {author} {\bibfnamefont {S.~L.}\ \bibnamefont {Rolston}}, \ and\
  \bibinfo {author} {\bibfnamefont {L.~A.}\ \bibnamefont {Orozco}},\ }\href
  {\doibase 10.1103/PhysRevA.78.013834} {\bibfield  {journal} {\bibinfo
  {journal} {Phys. Rev. A}\ }\textbf {\bibinfo {volume} {78}},\ \bibinfo
  {pages} {013834} (\bibinfo {year} {2008})}\BibitemShut {NoStop}%
\bibitem [{\citenamefont {Becerra}\ \emph {et~al.}(2010)\citenamefont
  {Becerra}, \citenamefont {Willis}, \citenamefont {Rolston}, \citenamefont
  {Carmichael},\ and\ \citenamefont {Orozco}}]{Becerra2010}%
  \BibitemOpen
  \bibfield  {author} {\bibinfo {author} {\bibfnamefont {F.~E.}\ \bibnamefont
  {Becerra}}, \bibinfo {author} {\bibfnamefont {R.~T.}\ \bibnamefont {Willis}},
  \bibinfo {author} {\bibfnamefont {S.~L.}\ \bibnamefont {Rolston}}, \bibinfo
  {author} {\bibfnamefont {H.~J.}\ \bibnamefont {Carmichael}}, \ and\ \bibinfo
  {author} {\bibfnamefont {L.~A.}\ \bibnamefont {Orozco}},\ }\href {\doibase
  10.1103/PhysRevA.82.043833} {\bibfield  {journal} {\bibinfo  {journal} {Phys.
  Rev. A}\ }\textbf {\bibinfo {volume} {82}},\ \bibinfo {pages} {043833}
  (\bibinfo {year} {2010})}\BibitemShut {NoStop}%
\bibitem [{\citenamefont {K\"{o}lle}\ \emph {et~al.}(2012)\citenamefont
  {K\"{o}lle}, \citenamefont {Epple}, \citenamefont {K\"{u}bler}, \citenamefont
  {L\"{o}w},\ and\ \citenamefont {Pfau}}]{Kolle2012}%
  \BibitemOpen
  \bibfield  {author} {\bibinfo {author} {\bibfnamefont {A.}~\bibnamefont
  {K\"{o}lle}}, \bibinfo {author} {\bibfnamefont {G.}~\bibnamefont {Epple}},
  \bibinfo {author} {\bibfnamefont {H.}~\bibnamefont {K\"{u}bler}}, \bibinfo
  {author} {\bibfnamefont {R.}~\bibnamefont {L\"{o}w}}, \ and\ \bibinfo
  {author} {\bibfnamefont {T.}~\bibnamefont {Pfau}},\ }\href {\doibase
  10.1103/PhysRevA.85.063821} {\bibfield  {journal} {\bibinfo  {journal} {Phys.
  Rev. A}\ }\textbf {\bibinfo {volume} {85}},\ \bibinfo {pages} {063821}
  (\bibinfo {year} {2012})}\BibitemShut {NoStop}%
\bibitem [{\citenamefont {Huber}\ \emph {et~al.}(2014)\citenamefont {Huber},
  \citenamefont {K\"{o}lle},\ and\ \citenamefont {Pfau}}]{Huber2014a}%
  \BibitemOpen
  \bibfield  {author} {\bibinfo {author} {\bibfnamefont {B.}~\bibnamefont
  {Huber}}, \bibinfo {author} {\bibfnamefont {A.}~\bibnamefont {K\"{o}lle}}, \
  and\ \bibinfo {author} {\bibfnamefont {T.}~\bibnamefont {Pfau}},\ }\href
  {\doibase 10.1103/PhysRevA.90.053806} {\bibfield  {journal} {\bibinfo
  {journal} {Phys. Rev. A}\ }\textbf {\bibinfo {volume} {90}},\ \bibinfo
  {pages} {053806} (\bibinfo {year} {2014})}\BibitemShut {NoStop}%
\bibitem [{\citenamefont {M\"{u}ller}\ \emph {et~al.}(2013)\citenamefont
  {M\"{u}ller}, \citenamefont {K\"{o}lle}, \citenamefont {L\"{o}w},
  \citenamefont {Pfau}, \citenamefont {Calarco},\ and\ \citenamefont
  {Montangero}}]{Muller2013}%
  \BibitemOpen
  \bibfield  {author} {\bibinfo {author} {\bibfnamefont {M.~M.}\ \bibnamefont
  {M\"{u}ller}}, \bibinfo {author} {\bibfnamefont {A.}~\bibnamefont
  {K\"{o}lle}}, \bibinfo {author} {\bibfnamefont {R.}~\bibnamefont {L\"{o}w}},
  \bibinfo {author} {\bibfnamefont {T.}~\bibnamefont {Pfau}}, \bibinfo {author}
  {\bibfnamefont {T.}~\bibnamefont {Calarco}}, \ and\ \bibinfo {author}
  {\bibfnamefont {S.}~\bibnamefont {Montangero}},\ }\href {\doibase
  10.1103/PhysRevA.87.053412} {\bibfield  {journal} {\bibinfo  {journal} {Phys.
  Rev. A}\ }\textbf {\bibinfo {volume} {87}},\ \bibinfo {pages} {053412}
  (\bibinfo {year} {2013})}\BibitemShut {NoStop}%
\bibitem [{\citenamefont {Willis}\ \emph {et~al.}(2011)\citenamefont {Willis},
  \citenamefont {Becerra}, \citenamefont {Orozco},\ and\ \citenamefont
  {Rolston}}]{Willis2011}%
  \BibitemOpen
  \bibfield  {author} {\bibinfo {author} {\bibfnamefont {R.~T.}\ \bibnamefont
  {Willis}}, \bibinfo {author} {\bibfnamefont {F.~E.}\ \bibnamefont {Becerra}},
  \bibinfo {author} {\bibfnamefont {L.~A.}\ \bibnamefont {Orozco}}, \ and\
  \bibinfo {author} {\bibfnamefont {S.~L.}\ \bibnamefont {Rolston}},\ }\href
  {\doibase 10.1364/OE.19.014632} {\bibfield  {journal} {\bibinfo  {journal}
  {Opt. Express}\ }\textbf {\bibinfo {volume} {19}},\ \bibinfo {pages} {14632}
  (\bibinfo {year} {2011})}\BibitemShut {NoStop}%
\bibitem [{\citenamefont {Zhang}\ \emph {et~al.}(2014)\citenamefont {Zhang},
  \citenamefont {Ding}, \citenamefont {Pan},\ and\ \citenamefont
  {Shi}}]{Zhang2014}%
  \BibitemOpen
  \bibfield  {author} {\bibinfo {author} {\bibfnamefont {W.}~\bibnamefont
  {Zhang}}, \bibinfo {author} {\bibfnamefont {D.-S.}\ \bibnamefont {Ding}},
  \bibinfo {author} {\bibfnamefont {J.-S.}\ \bibnamefont {Pan}}, \ and\
  \bibinfo {author} {\bibfnamefont {B.-S.}\ \bibnamefont {Shi}},\ }\href
  {\doibase 10.1088/0256-307X/31/6/064208} {\bibfield  {journal} {\bibinfo
  {journal} {Chinese Phys. Lett.}\ }\textbf {\bibinfo {volume} {31}},\ \bibinfo
  {pages} {064208} (\bibinfo {year} {2014})}\BibitemShut {NoStop}%
\bibitem [{\citenamefont {Srivathsan}\ \emph {et~al.}(2013)\citenamefont
  {Srivathsan}, \citenamefont {Gulati}, \citenamefont {Chng}, \citenamefont
  {Maslennikov}, \citenamefont {Matsukevich},\ and\ \citenamefont
  {Kurtsiefer}}]{Srivathsan2013a}%
  \BibitemOpen
  \bibfield  {author} {\bibinfo {author} {\bibfnamefont {B.}~\bibnamefont
  {Srivathsan}}, \bibinfo {author} {\bibfnamefont {G.~K.}\ \bibnamefont
  {Gulati}}, \bibinfo {author} {\bibfnamefont {B.}~\bibnamefont {Chng}},
  \bibinfo {author} {\bibfnamefont {G.}~\bibnamefont {Maslennikov}}, \bibinfo
  {author} {\bibfnamefont {D.}~\bibnamefont {Matsukevich}}, \ and\ \bibinfo
  {author} {\bibfnamefont {C.}~\bibnamefont {Kurtsiefer}},\ }\href {\doibase
  10.1103/PhysRevLett.111.123602} {\bibfield  {journal} {\bibinfo  {journal}
  {Phys. Rev. Lett.}\ }\textbf {\bibinfo {volume} {111}},\ \bibinfo {pages}
  {123602} (\bibinfo {year} {2013})}\BibitemShut {NoStop}%
\bibitem [{\citenamefont {Gulati}\ \emph {et~al.}(2014)\citenamefont {Gulati},
  \citenamefont {Srivathsan}, \citenamefont {Chng}, \citenamefont {Cer\`{e}},
  \citenamefont {Matsukevich},\ and\ \citenamefont {Kurtsiefer}}]{Gulati2014a}%
  \BibitemOpen
  \bibfield  {author} {\bibinfo {author} {\bibfnamefont {G.~K.}\ \bibnamefont
  {Gulati}}, \bibinfo {author} {\bibfnamefont {B.}~\bibnamefont {Srivathsan}},
  \bibinfo {author} {\bibfnamefont {B.}~\bibnamefont {Chng}}, \bibinfo {author}
  {\bibfnamefont {A.}~\bibnamefont {Cer\`{e}}}, \bibinfo {author}
  {\bibfnamefont {D.}~\bibnamefont {Matsukevich}}, \ and\ \bibinfo {author}
  {\bibfnamefont {C.}~\bibnamefont {Kurtsiefer}},\ }\href {\doibase
  10.1103/PhysRevA.90.033819} {\bibfield  {journal} {\bibinfo  {journal} {Phys.
  Rev. A}\ }\textbf {\bibinfo {volume} {90}},\ \bibinfo {pages} {033819}
  (\bibinfo {year} {2014})}\BibitemShut {NoStop}%
\bibitem [{\citenamefont {Autler}\ and\ \citenamefont
  {Townes}(1955)}]{Autler1955}%
  \BibitemOpen
  \bibfield  {author} {\bibinfo {author} {\bibfnamefont {S.~H.}\ \bibnamefont
  {Autler}}\ and\ \bibinfo {author} {\bibfnamefont {C.~H.}\ \bibnamefont
  {Townes}},\ }\href {\doibase 10.1103/PhysRev.100.703} {\bibfield  {journal}
  {\bibinfo  {journal} {Phys. Rev.}\ }\textbf {\bibinfo {volume} {100}},\
  \bibinfo {pages} {703} (\bibinfo {year} {1955})}\BibitemShut {NoStop}%
\bibitem [{\citenamefont {Wei}\ \emph {et~al.}(1998)\citenamefont {Wei},
  \citenamefont {Suter}, \citenamefont {Windsor},\ and\ \citenamefont
  {Manson}}]{Wei1998}%
  \BibitemOpen
  \bibfield  {author} {\bibinfo {author} {\bibfnamefont {C.}~\bibnamefont
  {Wei}}, \bibinfo {author} {\bibfnamefont {D.}~\bibnamefont {Suter}}, \bibinfo
  {author} {\bibfnamefont {A.~S.~M.}\ \bibnamefont {Windsor}}, \ and\ \bibinfo
  {author} {\bibfnamefont {N.~B.}\ \bibnamefont {Manson}},\ }\href {\doibase
  10.1103/PhysRevA.58.2310} {\bibfield  {journal} {\bibinfo  {journal} {Phys.
  Rev. A}\ }\textbf {\bibinfo {volume} {58}},\ \bibinfo {pages} {2310}
  (\bibinfo {year} {1998})}\BibitemShut {NoStop}%
\bibitem [{\citenamefont {Steck}(2013)}]{Steck2012}%
  \BibitemOpen
  \bibfield  {author} {\bibinfo {author} {\bibfnamefont {D.~A.}\ \bibnamefont
  {Steck}},\ }\href {http://steck.us/alkalidata/rubidium85numbers.pdf}
  {\enquote {\bibinfo {title} {{Rubidium 85 D line data}},}\ } (\bibinfo {year}
  {2013})\BibitemShut {NoStop}%
\bibitem [{\citenamefont {Happer}\ \emph {et~al.}(2010)\citenamefont {Happer},
  \citenamefont {Jau},\ and\ \citenamefont {Walker}}]{Happer2010}%
  \BibitemOpen
  \bibfield  {author} {\bibinfo {author} {\bibfnamefont {W.}~\bibnamefont
  {Happer}}, \bibinfo {author} {\bibfnamefont {Y.-Y.}\ \bibnamefont {Jau}}, \
  and\ \bibinfo {author} {\bibfnamefont {T.}~\bibnamefont {Walker}},\ }\href
  {\doibase 10.1002/9783527629503} {\emph {\bibinfo {title} {{Optically Pumped
  Atoms}}}}\ (\bibinfo  {publisher} {Wiley-VCH Verlag GmbH \& Co. KGaA},\
  \bibinfo {address} {Weinheim, Germany},\ \bibinfo {year} {2010})\BibitemShut
  {NoStop}%
\bibitem [{\citenamefont {Chrapkiewicz}\ and\ \citenamefont
  {Wasilewski}(2012)}]{Chrapkiewicz2012}%
  \BibitemOpen
  \bibfield  {author} {\bibinfo {author} {\bibfnamefont {R.}~\bibnamefont
  {Chrapkiewicz}}\ and\ \bibinfo {author} {\bibfnamefont {W.}~\bibnamefont
  {Wasilewski}},\ }\href {\doibase 10.1364/OE.20.029540} {\bibfield  {journal}
  {\bibinfo  {journal} {Opt. Express}\ }\textbf {\bibinfo {volume} {20}},\
  \bibinfo {pages} {29540} (\bibinfo {year} {2012})}\BibitemShut {NoStop}%
\bibitem [{\citenamefont {Parniak}\ and\ \citenamefont
  {Wasilewski}(2014)}]{Parniak2013a}%
  \BibitemOpen
  \bibfield  {author} {\bibinfo {author} {\bibfnamefont {M.}~\bibnamefont
  {Parniak}}\ and\ \bibinfo {author} {\bibfnamefont {W.}~\bibnamefont
  {Wasilewski}},\ }\href {\doibase 10.1007/s00340-013-5712-y} {\bibfield
  {journal} {\bibinfo  {journal} {Appl. Phys. B}\ }\textbf {\bibinfo {volume}
  {116}},\ \bibinfo {pages} {415} (\bibinfo {year} {2014})}\BibitemShut
  {NoStop}%
\bibitem [{\citenamefont {Balabas}\ \emph {et~al.}(2010)\citenamefont
  {Balabas}, \citenamefont {Jensen}, \citenamefont {Wasilewski}, \citenamefont
  {Krauter}, \citenamefont {Madsen}, \citenamefont {M\"{u}ller}, \citenamefont
  {Fernholz},\ and\ \citenamefont {Polzik}}]{Balabas2010}%
  \BibitemOpen
  \bibfield  {author} {\bibinfo {author} {\bibfnamefont {M.~V.}\ \bibnamefont
  {Balabas}}, \bibinfo {author} {\bibfnamefont {K.}~\bibnamefont {Jensen}},
  \bibinfo {author} {\bibfnamefont {W.}~\bibnamefont {Wasilewski}}, \bibinfo
  {author} {\bibfnamefont {H.}~\bibnamefont {Krauter}}, \bibinfo {author}
  {\bibfnamefont {L.~S.}\ \bibnamefont {Madsen}}, \bibinfo {author}
  {\bibfnamefont {J.~H.}\ \bibnamefont {M\"{u}ller}}, \bibinfo {author}
  {\bibfnamefont {T.}~\bibnamefont {Fernholz}}, \ and\ \bibinfo {author}
  {\bibfnamefont {E.~S.}\ \bibnamefont {Polzik}},\ }\href {\doibase
  10.1364/OE.18.005825} {\bibfield  {journal} {\bibinfo  {journal} {Opt.
  Express}\ }\textbf {\bibinfo {volume} {18}},\ \bibinfo {pages} {5825}
  (\bibinfo {year} {2010})}\BibitemShut {NoStop}%
\end{thebibliography}%

\end{document}